\documentclass{article}
\usepackage{calc,authblk,cite,braket}
\usepackage{tensor,float}
\usepackage{amssymb}
\usepackage{mathrsfs}
\usepackage{mciteplus}
\usepackage{geometry}
\usepackage{amsmath}
\usepackage{blindtext}
 \usepackage{braket}
\usepackage{epigraph}
\usepackage{tikz}
\usepackage{mathalpha}
\usetikzlibrary{arrows.meta}
\usepackage{blindtext}
\usepackage{pdfpages}
\usepackage{tabularx}
\usepackage{etoolbox}
\AtEndEnvironment{figure}{\par\indent}
\AtEndEnvironment{table}{\par\indent}
\usepackage{graphicx}
\usepackage{caption}
\usepackage{subcaption}
\newcommand{\Erf}{\ensuremath{\operatorname{Erf}}}

\title{Quantum Correlations and Gravity: From the Emergence of a Cosmological Constant to the Gravitation of Particles in Superposition}
\author{Johas Morales and Yuri Bonder$^*$}
\affil{$^*$bonder@nucleares.unam.mx\\
Instituto de Ciencias Nucleares\\
Universidad Nacional Aut\'onoma de M\'exico\\
Circuito Exterior s/n, Ciudad Universitaria, Cd.Mx., 04510, M\'exico}

\begin{document}

\maketitle

\begin{abstract}
One of the main technical obstacles in constructing a consistent theory of quantum gravity is that the metric itself defines the causal structure required for quantization. This motivates implementing quantum aspects of gravity through an independent connection. Moreover, the experimentally confirmed violation of Bell inequalities, together with the natural structure of the energy--momentum tensor in semiclassical gravity, suggests that nonlocality should be incorporated into the gravitational formalism. Motivated by these considerations, we propose a model in which the connection is treated as an independent bitensorial field, leading to a bitensorial generalization of the Einstein equations. The model reduces to General Relativity when the matter source is classical. We apply it in two regimes: the late-time universe and the Newtonian limit. In the cosmological case, the model naturally gives rise to a positive effective cosmological constant. In the Newtonian regime, we analyze a situation in which the gravitational source is in a quantum superposition and find that the model predicts a novel, nonconservative effective force that depends on the velocity of the test particle.
\end{abstract}

\section{Introduction}

General Relativity (GR) is our current theory of gravitation, where gravity is the result of the curvature of spacetime. The theory has achieved remarkable empirical success~\cite{Will_2014}, accounting for phenomena such as gravitational redshift~\cite{Bothwell2022}, the Hulse–Taylor binary pulsar~\cite{HulseTaylorBinary}, frame dragging~\cite{gravityprobeb}, and gravitational waves~\cite{ondas}. Nevertheless, GR faces several fundamental challenges. These include the need for dark matter and dark energy~\cite{energiaoscura,Debono_2016}, the appearance of spacetime singularities~\cite{Penrose}, and its incompatibility with quantum mechanics~\cite{Carlip}. These issues strongly suggest that GR is not the most fundamental description of gravity.

Quantum mechanics, by contrast, is built on an entirely different conceptual framework~\cite{Sakurai}. Its kinematics rely on Hilbert spaces, with physical quantities represented by operators, which in turn allows these quantities to acquire inherently ``fuzzy'' values. While quantum theory also enjoys extraordinary experimental success~\cite{Leggett_2002,Fan_130}, it leaves key conceptual questions unresolved. The theory postulates a collapse of the wave function during measurements, without clearly defining what constitutes a measurement~\cite{Maudlin1995ThreeMP}, and its ontology remains disputed~\cite{Albert2013-WaveFunctionRealism,Goldstein2013-RealityRoleWaveFunction,Maudlin2007-CompletenessSupervenienceOntology}. 

Reconciling GR and quantum mechanics into a consistent theory of quantum gravity remains one of the major open problems in theoretical physics~\cite{Carlip_2001}. Moreover, the absence of experimental guidance makes the task of constructing such a theory a formidable one. Thus, any empirical indication of how a quantum system gravitates is invaluable, as illustrated by the attention that proposals to detect gravity-induced entanglement have attracted~\cite{MarlettoVedral,Bose}.

The problem of quantum gravity is \emph{not} merely a matter of unification~\cite{Weinberg1993,PhysRevX.13.041040}, renormalization~\cite{Boulware:1974cn,BERN1998401,Horava}, or generating a discrete description of spacetime~\cite{Snyder,Sorkin,Bonder_2008,PerezSudarsky}. The issue arises from the structure of Einstein’s field equations, which couple geometry and matter through\footnote{We employ Planck units with \( c = \hbar = G = 1 \). Lowercase Latin letters denote abstract tensor indices. We work in four spacetime dimensions and the metric signature is \( (-+++) \). Other conventions follow Ref.~\cite{Wald1984}.}
\begin{equation}\label{1}
G_{ab}(g) = 8\pi\, T_{ab}(\hat{\phi}),
\end{equation}
where the Einstein tensor, \( G_{ab} \), depends on the spacetime metric, \( g_{ab} \), while the energy–momentum tensor, \( T_{ab} \), describes the matter fields, here represented by quantum operators \( \hat{\phi} \). The problem is that quantum matter lacks definite classical properties, such as position or energy, making it unclear how the spacetime geometry should respond.

Perhaps the best approach to this problem is semiclassical gravity~\cite{Moller1962,Rosenfeld1963,Ford,Wald1994,Decanini_2008}, where Eq.~\eqref{1} is replaced by
\begin{equation}
G_{ab}(g) = 8\pi \langle T_{ab}(\hat{\phi}) \rangle, \label{semiclassical}
\end{equation}
so that the right-hand side incorporates quantum effects through expectation values. Constructing these expectation values, however, is technically cumbersome \cite{Decanini_2008}. For example, in the case of a free, massive, minimally coupled Klein–Gordon field,
\begin{equation}
T_{ab}(\hat{\phi}) = \nabla_a \hat{\phi}\, \nabla_b \hat{\phi} 
	- \frac{1}{2} g_{ab} \left( \nabla_c \hat{\phi}\, \nabla^c \hat{\phi} + m^2 \hat{\phi}^2 \right),
\end{equation}
which is quadratic in \( \hat{\phi} \). However, according to quantum fields on curved spacetime~\cite{Parker1969a,Parker1971,BD1982,fulling,Wald1994,parkerQFT}, \( \hat{\phi} \) are operator-valued distributions and products such as \( \hat{\phi}^2 \) are ill-defined.

A standard approach around this issues is point-splitting regularization~\cite{Christensen_1976}, which, for the Klein–Gordon field, introduces a two-point operator
\begin{equation}
\hat{T}_{ab'}(x,x') = \nabla_a \nabla_{b'} 
	- \frac{1}{2} g_{ab'} \left( \nabla_c \nabla^{c'} + m^2 \right).
\end{equation}
With this operator, one defines the expectation value of the energy--momentum tensor as
\begin{equation}
\langle T_{ab}(\hat{\phi}) \rangle
	= \lim_{x' \to x} 
	\hat{T}_{ab'}(x,x')
	\langle \psi | \hat{\phi}(x)\hat{\phi}(x') | \psi \rangle_{\text{ren}},
\end{equation}
where \( \langle \psi | \hat{\phi}(x)\hat{\phi}(x') | \psi \rangle_{\text{ren}} \) is the regularized (and renormalized) two-point function. To regularize the two-point function, one must restrict the admissible states to the Hadamard class~\cite{Decanini_2008}, whose short-distance behavior coincides with that of flat spacetime~\cite{Decanini_2008}, which is known and can thus be subtracted. Once this function is regularized, the coincidence limit can be taken to restore locality. However, this procedure relies on a technically unnatural regularization and a restrictive choice of states. This observation motivates us to search for formulations that avoid taking the coincidence limit altogether. Furthermore, semiclassical gravity faces other conceptual difficulties: the left-hand side of Eq.~\eqref{semiclassical} is divergence-free by construction, whereas quantum collapse is ``instantaneous.'' As a result, during measurement processes, the right-hand side cannot be divergence-free,  signaling a breakdown of the semiclassical description~\cite{Adolfo,Diez}.

At a more fundamental level, the tension between gravity and quantum theory can be viewed in terms of microcausality \cite{weinbergQFT}. In quantum field theory, local operators at spacelike separation (anti)commute, ensuring consistency with the causal spacetime structure. Quantizing gravity, however, entails quantizing the metric itself, which is the object that defines spacelike separation. This circularity is at the root of the challenges faced in attempts to quantize gravity.

A possible way forward is to reconsider the geometrical foundations of GR. In Einstein’s formulation, the metric is the sole dynamical variable. Yet, more than a century ago, Palatini~\cite{Palatini1919} demonstrated that treating the affine connection \( k^{a}_{\ bc} \) as an independent variable in the Einstein–Hilbert action leads to equations dynamically equivalent to those of GR (provided that the matter action is independent of the connection~\cite{italianos}). This motivates us to explore extensions in which the connection acquires quantum properties. One may thus envisage an equation of the form
\begin{equation}
G_{ab}(g,\hat{k}) = 8\pi T_{ab}(\hat{\phi}),
\end{equation}
where the metric remains classical, preserving the causal structure, while the connection incorporates the quantum effects.

Constructing a complete theory of this kind remains out of reach. Our aim, therefore, is to introduce into the gravitational description, while staying within the geometrical framework, possible phenomenological modifications in which the connection encodes an experimentally established quantum feature. We propose that the key property to incorporate is nonlocality, which emerges when the coincidence limit is not taken and becomes manifest in Bell-type correlations.

More concretely, we seek to formulate a theory in which the right-hand side of Einstein’s equations is given by the expectation value of an energy–-momentum tensor without taking the coincidence limit. This requires introducing, on the left-hand side, an Einstein-like bitensor. The corresponding field equations take the form
\begin{equation}\label{EinsteinBitensorial}
G_{ab'}(x, x') = 8 \pi \braket{T_{ab'}(x, x')}.
\end{equation}
In our approach, $G_{ab'}(x, x')$ is constructed from an independent, bitensorial connection, the latter encoding the nonlocal aspects. The next section is devoted to introducing the conceptual and technical tools required for the development of the rest of the paper.

\section{Preliminaries}

\subsection{Bell's Theorem}

Starting with Bell’s seminal work~\cite{belloriginal}, an empirically testable criterion was established to distinguish between theories that respect the relativistic causal structure and those that do not. In its current formulation, this criterion takes the form of an inequality whose violation signals the presence of effects incompatible with relativistic causality~\cite{beables,GHZ}. Experimental tests of Bell’s inequalities began in the 1980s, and it is now an empirical fact that these inequalities are violated~\cite{bellexperimento,bellexperimento2,bellexperimento3}. Importantly, a defining feature of Bell’s analysis is its theory-agnostic character: it does not rely on the postulates of quantum mechanics, even though quantum theory does predict such violations.

The derivation of Bell’s inequalities rests on two assumptions~\cite{hipotesisbell,maudlinnonlocality}. The first is \emph{locality}, which states that what occurs at a spacetime event depends only on its past light cone. The second is \emph{settings independence}, which asserts that experimenters’ choices of the settings of a measurement are not predetermined. This latter assumption is a cornerstone of the scientific method; rejecting it would undermine the very foundations of empirical science. Faced with the conclusion that at least one of Bell’s assumptions is false, we interpret the situation as indicating that nature exhibits nonlocal features (albeit ones that do not allow faster-than-light communication~\cite{Casella2023-CASCAN}).

These nonlocal aspects reveal a fundamental tension with GR, in which only tensorial fields propagating according to the spacetime causal structure are admissible. Nevertheless, any attempt to construct a gravitational theory that is less incompatible with quantum mechanics must address this tension. In this work, we propose a theory with these features, after introducing the necessary mathematical tools in the following section. We now turn to a brief review of some basic properties of an affine connection.

\subsection{Affine connection}

An affine connection $\tensor{C}{^c_{ab}}$ specifies how vectors are parallel transported along curves in spacetime. For brevity, we refer to such tensor fields simply as connections. These tensors define an associated derivative operator~\cite{Wald1984}, $\nabla_a$, which we take to be metric compatible, namely,
\begin{equation}
\nabla_a g_{bc} = 0.
\end{equation}

The commutator of two covariant derivatives is related to infinitesimal holonomies, and, when acting on an arbitrary vector field, takes the form\footnote{Parentheses (square brackets) denote (anti)symmetrization over the enclosed indices, including a factor of $1/n!$ for $n$ indices, while indices placed between vertical bars are excluded from the (anti)symmetrization.}
\begin{equation}\label{antisimetrizacionderivadas}
2\nabla_{[a}\nabla_{b]}u^c = \tensor{R}{^c_{d a b}} u^d + 2\tensor{S}{_{ab}^d}\nabla_d u^c,
\end{equation}
where the curvature tensor, $\tensor{R}{^d_{c a b}}$, can be expressed as 
\begin{equation}\label{Riemannconexion}
\tensor{R}{^d_{c a b}} = 2\partial_{[a}\tensor{C}{^d_{|c|b]}} + 2\tensor{C}{^d_{e [a}}\tensor{C}{^e_{|c|b]}},
\end{equation}
and the torsion tensor is defined by 
\begin{equation}
\tensor{S}{_{ab}^c} = \tensor{C}{^c_{[ab]}}.
\end{equation}

The Ricci tensor is constructed by taking a contraction of the curvature tensor, namely,
\begin{equation}
\tensor{R}{_a_b}=\tensor{R}{^c_a_c_b}.
\end{equation}
By contracting the Ricci tensor with the inverse metric, one obtains the Ricci scalar,
\begin{equation}
R=R_{ab}g^{ab}.
\end{equation}
Finally, the Einstein tensor is defined as
\begin{equation}
G_{ab}=R_{ab}-\frac{1}{2}g_{ab}R.
\end{equation}

A metric defines a unique torsion-free connection given by
\begin{equation}
\tensor{\mathring{\Gamma}}{^c_{ab}} = \frac{1}{2} g^{cd} \left( \partial_a g_{bd} + \partial_b g_{ad} - \partial_d g_{ab} \right),
\end{equation}
and whose associated derivative operator is $\mathring{\nabla}_a$. In general, a connection can be decomposed into a metric part and a torsion part as
\begin{equation}\label{descomposicionconexion}
\tensor{C}{^c_{ab}} = \tensor{\mathring{\Gamma}}{^c_{ab}} + \tensor{k}{_{ab}^c},
\end{equation}
where\footnote{Spacetime indices are lowered and raised using the metric and its inverse, $g^{ab}$, as is customary.}
\begin{equation}
\tensor{k}{_{ab}^c} = g^{cd}\left(S_{dab} + S_{dba} - S_{abd}\right)
\end{equation}
is the contorsion tensor, which satisfies $\tensor{k}{_{abc}} = -\tensor{k}{_{cba}}$. Consequently, the contorsion tensor has $24$ independent components.  

The commutator of $\mathring{\nabla}_a$ defines another curvature tensor, $\tensor{\mathring{R}}{^d_{c a b}}$, which is related to $\tensor{R}{^d_{c a b}}$ through
\begin{equation}
\tensor{R}{^d_{c a b}} =
\tensor{\mathring{R}}{^d_{c a b}} 
- \mathring{\nabla}_a \tensor{k}{_{c b}^d} 
+ \mathring{\nabla}_b \tensor{k}{_{c a}^d}
+ \tensor{k}{_{e a}^d} \tensor{k}{_{c b}^e} 
- \tensor{k}{_{e b}^d} \tensor{k}{_{c a}^e}.
\end{equation}
It is also possible to define an Einstein tensor associated with $\mathring{\nabla}_a$, denoted by $\mathring{G}_{ab}$, which satisfies
\begin{align}\nonumber
G_{ab} = \, & \mathring{G}_{ab}
- \mathring{\nabla}_d \tensor{k}{_{a b}^d}
+ \mathring{\nabla}_b \tensor{k}{_{a d}^d}
+ \tensor{k}{_{e d}^d} \tensor{k}{_{a b}^e}
- \tensor{k}{_{e b}^d} \tensor{k}{_{a d}^e} \\
\label{EinsteinK}
& - \frac{1}{2} g_{ab} g^{ef} 
\left(
- \mathring{\nabla}_d \tensor{k}{_{e f}^d}
+ \mathring{\nabla}_f \tensor{k}{_{d e}^d}
+ \tensor{k}{_{s d}^d} \tensor{k}{_{e f}^h}
- \tensor{k}{_{h f}^d} \tensor{k}{_{e d}^h}
\right).
\end{align}
Notice that, in general, $G_{ab} \neq G_{ba}$. In the next subsection we introduce the notion of bitensors.

\subsection{Bitensors}

A bitensor is a mathematical object, introduced long ago~\cite{DEWITT}, that takes $k$ covectors and $\ell$ vectors at a point $x$, and $k'$ covectors and $\ell'$ vectors at a point $x'$, yielding a real number in a multilinear fashion. Such bitensors are denoted by
\begin{equation}\label{klk'l'}
\tensor{T}{^{a_{1}\ldots a_{k}}_{b_{1}\ldots b_{\ell}}^{a'_{1}\ldots a'_{k'}}_{b'_{1}\ldots b'_{\ell'}}},
\end{equation}
where unprimed indices refer to vectors and covectors at $x$, and primed indices to those at $x'$. A bitensor that accepts the above number of arguments is said to be of type $(k,\ell)\times(k',\ell')$.

As becomes clear below, the bitensors we use are only well defined within a convex normal neighborhood, that is, an open subset of spacetime such that any pair of points in it can be connected by a unique geodesic entirely contained in this subset. In general, this condition does not hold globally. Hence, we restrict ourselves, when necessary, to convex normal neighborhoods by introducing suitable cutoffs.

We now introduce some well-known bitensors. A relevant example is the parallel transporter,
\begin{equation}
\tensor{g}{^{a'}_b}=\tensor{g}{^{a'}_b}(x,x'),
\end{equation}
which encodes the operation of parallel transport between the tangent spaces at $x$ and $x'$ along the unique geodesic connecting them. When it acts on a vector $v^a$ at $x$, it yields its parallel transported counterpart at $x'$,
\begin{equation}
v^{a'}=\tensor{g}{^{a'}_b}v^b.
\end{equation}
Note that \(x'\) can be arbitrarily far from \(x\). Thus, determining the parallel transporter requires integrating the connection along a path, since the connection only provides the rule for parallel transport between infinitesimally close points.

The inverse operation, namely parallel transport from $x'$ to $x$, is represented by
\begin{equation}
\tensor{g}{^a_{b'}}=\tensor{g}{^a_{b'}}(x,x').
\end{equation}
In general, the composition of parallel transporters depends on the intermediate point,
\begin{equation}
\tensor{g}{^{a''}_{b'}}\tensor{g}{^{b'}_c}
  \neq
\tensor{g}{^{a''}_{b'''}}\tensor{g}{^{b'''}_c},
\end{equation}
where double and triple primed indices refer to two arbitrary and distinct points, $x''$ and $x'''$, respectively. This property reflects the presence of curvature.

The parallel transporter also allows one to construct bitensors from ordinary tensors. For instance,
\begin{equation}
\tensor{T}{^{a'}_b}=\tensor{g}{^{a'}_a}\tensor{T}{^a_b}.
\end{equation}
Finally, one may define a parallel transporter along autoparallel curves~\cite{Syngenogeodesico}, that is, curves whose tangent vectors are parallel transported along themselves and which, in general, do not coincide with geodesics.

An intrinsic operation on bitensors is the coincidence limit $x' \to x$. This limit does not change the total number of covariant and contravariant indices of the bitensor: for a bitensor of type $(k,\ell)\times(k',\ell')$, the coincidence limit yields a tensor of type $(k+k',\ell+\ell')$. A common notation for this operation is to enclose the bitensor in square brackets. For instance, the coincidence limit of the parallel transporter is
\begin{equation}
\left[\tensor{g}{^{a'}_b}\right]=\lim_{x' \to x}\tensor{g}{^{a'}_b}=\delta^a_b.
\end{equation}

Another well-known bitensor is Synge's world function~\cite{Poisson}, a bitensor of type $(0,0)\times(0,0)$ (also known as a bifunction) defined as
\begin{equation}
\sigma(x,x')=\frac{1}{2}\int_\gamma g_{ab}t^a t^b \, \text{d}\lambda,
\end{equation}
where $\gamma$ is the geodesic connecting $x$ and $x'$, affinely parametrized by $\lambda$, and $t^a$ is its tangent vector. Clearly,
\begin{equation}
\sigma=\frac{1}{2}\mu^2,
\end{equation}
where the geodesic distance is
\begin{equation}
\mu(x,x')=\int_\gamma \sqrt{g_{ab}t^a t^b} \, \text{d}\lambda.
\end{equation}
Taking the coincidence limit of Synge's world function gives $[\sigma]=0$.

The derivatives of bitensors are defined with respect to one of their arguments, keeping the other fixed. The following limits hold:
\begin{equation}
[\nabla_b\nabla_a\sigma] = g_{ab}, \qquad
[\nabla_{b'}\nabla_a\sigma] = -g_{ab}.
\end{equation}
These relations, which have been used to argue that \(\sigma\) may be an even more fundamental object than the metric~\cite{achim}, can be established in a normal coordinate system centered at \(x\). These are the mathematical tools required for our analysis, we now proceed to present the model.

\section{The Model}

To construct the proposed model, we first discuss the index structure of a bitensorial connection. We emphasize that the connection is a natural candidate for exhibiting nonlocal behavior, as it is the object that governs the transport of tensors between infinitesimally close points; an intrinsically nonlocal operation. However, the bitensorial connection introduced here relates tensors at points that may lie arbitrarily far apart.

The most natural geometrical interpretation of a bitensorial connection is as a bitensor 
\begin{equation}\label{abc'}
\tensor{k}{_a_b^{c'}}(x,x'),
\end{equation}
of type $(1,0)\times(0,2)$, which takes a vector $v^a$ at $x$ and the tangent to the geodesic linking $x$ and $x'$, also at $x$, $t^b$, and yields
\begin{equation}\label{naivek}
v^{c'}=\tensor{k}{_a_b^{c'}}(x,x')v^a t^b,
\end{equation}
which is a vector in $x'$. However, in general, this operation is not linear in $t^b$, despite what is suggested by a naive inspection of the right-hand side of Eq.~\eqref{naivek}. Therefore, we instead consider a bitensorial connection of type $(1,2)\times(0,0)$, related to that in Eq.~\eqref{naivek} by
\begin{equation}
\tensor{k}{_a_b^c}(x,x') = \tensor{g}{^c_{c'}} \tensor{k}{_a_b^{c'}}(x,x').
\end{equation}
This is the bitensorial connection that serves as the dynamical object in our proposal.

The Einstein bitensor that couples to the bitensorial energy--momentum tensor through Eq.~\eqref{EinsteinBitensorial} is\footnote{This model constitutes an improved version of our initial proposal~\cite{paper}.}
\begin{equation}\label{Ecuacionprincipalmodelo}
G_{ab'}(x,x') = \frac{1}{2}\Big[ \tensor{g}{^{b}_{b'}} G_{ab}
    + \tensor{g}{^{a'}_{a}} G_{a'b'} \Big],
\end{equation}
where, inspired by Eq.~\eqref{EinsteinK}, we propose
\begin{eqnarray}\label{Ecuacionsecundariamodelo}
G_{ab}(x,x') &=& \mathring{G}_{ab}
    + H_{ab}
    + \left(-\frac{1}{2} g_{ab} H_{cd} g^{cd}\right),\\[4pt]
\label{H}
H_{ab}(x,x') &=& - \mathring{\nabla}_c \tensor{k}{_{(ab)}^{c}}
    + \mathring{\nabla}_{(a} \tensor{k}{_{b)c}^{c}}
    + \tensor{k}{_{dc}^{c}} \tensor{k}{_{(ab)}^{d}}
    - \tensor{k}{_{d(a}^{c}} \tensor{k}{_{b)c}^{d}}.
\end{eqnarray}
In the last expression, $\tensor{k}{_{ab}^c} = \tensor{k}{_{ab}^c}(x,x')$ denotes the bitensorial connection, which is the dynamical quantity to be determined. Observe that in Eq.~\eqref{Ecuacionprincipalmodelo} the arguments of the first term are $(x,x')$, whereas in the second they appear reversed, $(x',x)$. Hence, the Einstein bitensor treats both points, $x$ and $x'$, on an equal footing.

Let us now consider the description of matter. Suppose that part of the matter content is in a ``classical'' or ``coherent'' state, which couples in the usual way to the Einstein tensor, while the remaining matter is in a ``quantum'' state. Then, assuming that the quantum part is well described by bitensors, the expectation value of the energy--momentum tensor can be decomposed as
\begin{equation}\label{formaTab}
\braket{T_{ab'}}=\frac{1}{2}\left(T_{ab}\tensor{g}{^b_{b'}} + T_{a'b'}\tensor{g}{^{a'}_a}\right) + \Delta T_{ab'},
\end{equation}
where $T_{ab}$ is a conventional tensor that describes the classical component, and $\Delta T_{ab'}$ is a bitensorial object, as suggested by its indices. Again, in the parenthesis on the right-hand side of Eq.~\eqref{formaTab}, the first term depends on $(x,x')$ and the second term depends on the points in reversed order. Then, the model naturally splits into a tensorial part and a bitensorial one, namely,
\begin{equation}
\mathring{G}_{ab}(x)=8 \pi T_{ab}(x),\label{Einstein}
\end{equation}
and
\begin{equation}
g^b_{b'}H_{ab}+g^{a'}_{a}H_{a'b'}-\frac{1}{2}(g^{cd}H_{cd}+g^{c'd'}H_{c'd'})g_{ab'}=16 \pi \Delta T_{ab'}(x,x').\label{BitensorialPart}
\end{equation}
Notice that Eq.~\eqref{Einstein} is the conventional Einstein equation, from which the metric can be obtained. Then, one can use the metric to find the parallel transporter required in the bitensorial part, Eq.~\eqref{BitensorialPart}. Equation~\eqref{BitensorialPart} can then be used to solve for the bitensorial connection $\tensor{k}{_a_b^c}(x,x')$.

Moreover, when all matter is classical, $\Delta T_{ab'}=0$, which implies $H_{ab}=0$ and, consequently, the vanishing of the bitensorial connection. In this limit, the model reduces to conventional GR. This feature is central to the proposal: the new geometric effects emerge exclusively in the presence of sources of gravity that are in ``quantum states,'' leaving the physics unaltered in the classical regime.

Several remarks are in order. First, note that the model yields two conventional Einstein equations, one for each point. However, since these are field equations, they are equivalent. Second, and more importantly, the equation for the bitensorial connection is underdetermined, as it provides $10$ equations for the $24$ components of the bitensorial connection. In this sense, the model clearly requires an additional equation. This situation is not new, as it mirrors what happens in theories with an independent connection, which are supplemented by an extra equation. In the most natural case, that of Einstein--Cartan theory~\cite{Hehl_1976,Hehl}, this equation is algebraic. Third, $H_{ab}$ is taken to be symmetric by construction. We believe that the last two issues are related and could be addressed without major difficulty. One may, for instance, introduce an equation for the connection analogous to that in Einstein--Cartan theory, and allow \(H_{ab}\) to possess a nontrivial antisymmetric part. Nevertheless, the present work constitutes a first exploration of a new class of theories, and to keep the technical aspects under control at this stage, we leave such refinements for future work.

In addition, note that the bitensorial contribution to the energy--momentum tensor, $\Delta T_{ab'}$, still diverges in the coincidence limit. Our proposal is to treat this divergence as a boundary condition for the bitensorial Einstein tensor. More concretely, we construct the latter so that it exhibits the same divergence behavior as the matter term when the two points coincide, thereby ensuring that the resulting field equations remain well defined even in this limit.

At this stage, the model is completely formulated and can, in principle, be solved for any bitensorial matter source \(\Delta T_{ab'}\). However, it is useful to introduce additional simplifying assumptions, which we discuss next.

\subsection{Perturbative scheme}\label{pert}

To further simplify the technical aspects of the model, we consider perturbations around Minkowski spacetime, which are suitable for describing either a small portion of the universe's history or a laboratory experiment. We also assume that the bitensorial contribution to the energy--momentum tensor is small compared with its classical part, since such effects have not yet been detected. In particular, we neglect quadratic terms in the metric perturbation and in the bitensorial part of the energy--momentum tensor, as well as terms involving products of these two quantities.

To formalize this approximation, we introduce a bookkeeping parameter $\epsilon$, such that $0<\epsilon \ll 1$, and discard all terms quadratic in this parameter. Accordingly, we write
\begin{eqnarray}   \label{metricaperturbativa}
g_{ab} &=& \eta_{ab} + \epsilon\, h_{ab} + \mathcal{O}(\epsilon^2),\\
\braket{T_{ab'}} &=& \frac{1}{2}\left( T_{ab}\tensor{g}{^b_{b'}} + T_{a'b'}\tensor{g}{^{a'}_a} \right)
+ \epsilon\, \Delta T_{ab'} + \mathcal{O}(\epsilon^2),\label{expansionTab}
\end{eqnarray}
where $\eta_{ab}$ denotes the flat spacetime metric. Clearly, Eq.~\eqref{expansionTab} requires us to implement a similar expansion in the left-hand side of the bitensorial Einstein equation, namely,
\begin{equation}\label{EcuacionsecundariamodeloPert}
G_{ab}(x,x')=\mathring{G}_{ab}+\epsilon\left(H_{ab}-\frac{1}{2}g_{ab}H_{cd}g^{cd}\right)+ \mathcal{O}(\epsilon^2).
\end{equation}
Hence, in this approximation, the bitensorial part of the Einstein equation is only sensitive to the flat metric. In particular, the parallel transporters trivialize in the sense that
$\tensor{g}{^{a'}_a} = \tensor{\delta}{^{a'}_a} + \mathcal{O}(\epsilon)$. Here the first term represents the identity operator, which, in global Lorentzian coordinates, adopted from now on, simply copies the components at $x$ to $x'$.

Under this approximation the linear part in $\epsilon$ of the dynamical equations, Eq.~\eqref{BitensorialPart}, becomes
\begin{equation}
\tensor{\delta}{^b_{b'}}\left(\chi_{ab} + \Xi_{ab}\right)
+\tensor{\delta}{^{a'}_a}\left(\chi_{a'b'} + \Xi_{a'b'}\right)
= 8 \pi\, \Delta T_{ab'}(x,x'),
\label{esquema1}
\end{equation}
where
\begin{eqnarray}
\chi_{ab} &=& -\partial_c\tensor{k}{_{(ab)}^c}
+ \partial_{(a}\tensor{k}{_{b)c}^c}
+ \frac{1}{2}\eta_{ab}\eta^{de}\left(\partial_c\tensor{k}{_d_e^c}
- \partial_d\tensor{k}{_e_c^c}\right),\\
\label{CuadraticosK}
\Xi_{ab} &=& \tensor{k}{_{e d} ^d} \tensor{k}{_{(ab)} ^ e}
            - \tensor{k}{_{e(a} ^d} \tensor{k}{_{b)d} ^ e}
            + \frac{1}{2} \eta_{ab} \eta^{de} \left[
            -\tensor{k}{_{f g} ^ g} \tensor{k}{_{d e} ^f}
            + \tensor{k}{_{g e} ^ f} \tensor{k}{_{d f} ^ g}
            \right].
\end{eqnarray}
Note that in $\chi_{a'b'}$ and $\Xi_{a'b'}$, all dummy indices are also primed and the arguments $x$ and $x'$ are reversed with respect to $\chi_{ab}$ and $\Xi_{ab}$.

When describing the effects of the model in a table-top experiment, one can assume that the bitensorial connection is itself small. This allow us to also neglect quadratic terms in the bitensorial connection, further simplfying the calculations. Under this expansion Eq.~\eqref{BitensorialPart} becomes
\begin{equation}
\tensor{\delta}{^b_{b'}} \chi_{ab}
+\tensor{\delta}{^{a'}_a} \chi_{a'b'}
= 8 \pi\, \Delta T_{ab'}(x,x').
\label{esquema2}
\end{equation}
Unfortunately, this approximation trivializes the effects in the cosmological setup. These two cases, the cosmological setting and the table-top experiment, are discussed in the following sections.

\section{Late-time Cosmology}\label{Sec:Cosmology}

In this section we analyze the behavior of the model in the late-time universe under the assumption of homogeneity and isotropy. Our main goal is to determine whether the model leads to observable consequences and, in particular, whether it can shed light on the origin of the cosmological constant, one of the most persistent puzzles in modern theoretical physics~\cite{weinbergconstante}.

We explore a universe filled with conventional dust, radiation, or a cosmological constant, described by the classical stress-energy tensor $T_{ab}$, while additional quantum matter effects are encoded in $\Delta T_{ab'}$. The cosmological time direction is aligned with the four-velocity $u^a$ of the comoving (classical) matter distribution. The coordinate $t$ parametrizes the integral curves of this vector field, i.e., $u^a = (\partial / \partial t)^a$.

The conventional part of the model, which is given by the conventional Einstein equation, allows us to consider that a Friedmann–Robertson–Walker (FRW) metric. For simplicity, we take this metric to be spatially flat. Thus, in suitable coordinates, the metric takes the form
\begin{equation}\label{cosmologicalmetric}
\text{d}s^2 = -\text{d}t^2 + a(t)^2 (\text{d}x^2 + \text{d}y^2 + \text{d}z^2).
\end{equation}
At late times near the present cosmological time $t_0$, the scale factor can be approximated as~\cite{expansiona}
\begin{equation}\label{expansionaa}
a(t) = 1 + \epsilon H_0 (t - t_0)  + \mathcal{O}(\epsilon^2),
\end{equation}
where we use the normalization $a(t_0) = 1$. Here, $\dot{a}(t_0) = H_0$ corresponds to the current value of the Hubble parameter, and overdots denote derivatives with respect to $t$. We also extract an expansion parameter $\epsilon$ from $H_0 (t - t_0)$, as in Subsec.~\ref{pert}.

Given the symmetries at hand, the most general form of the bitensorial energy--momentum tensor is 
\begin{equation}\label{sourcecosmo}
\Delta T_{ab'} = \rho(t,t')\,u_a u_{b'} + P(t,t')\,\delta^{(3)}_{ab'} + \mathcal{O}(\epsilon),
\end{equation}
where $\delta^{(3)}_{ab'} = \delta^{(3)}_{ab}\,\tensor{\delta}{^{b}_{b'}} + \mathcal{O}(\epsilon)$, and $\delta^{(3)}_{ab}$ is the ``flat part'' of the induced metric on the comoving hypersurfaces. The quantities $\rho$ and $P$ are bifunctions of $t$ and $t'$, symmetric under $t \leftrightarrow t'$, as required by the interchange symmetry of the model. An ansatz for the bitensorial connection is\footnote{We could include a term $k_{abc} \propto \varepsilon_{abc}$, where $\varepsilon_{abc}$ is the volume element of the comoving hypersurfaces. Such a term, however, is not invariant under spatial inversions. Therefore, we do not consider it.}
\begin{equation}\label{ansatzK}
\tensor{k}{_{abc}} = 2 u_{[a} \delta^{(3)}_{c]b}\, F(t,t'),
\end{equation}
where $F(t,t')=F(t',t)$, by construction.

We can readily verify that
\begin{eqnarray}
\partial_{d} \tensor{k}{_{(ab)}^d} &=&  \left[u_{(a} {\delta^{(3)}}^d_{b)} 
- \delta^{(3)}_{a b} u^d\right]\partial_d F(t, t')=- \delta^{(3)}_{ab}\frac{\partial F}{\partial t},\\
\partial_{(a}\tensor{k}{_{b)}_d^d} &=& 3u_{(a}\partial_{b)}F(t,t')= -3u_a u_b \frac{\partial F}{\partial t},\\
\tensor{k}{_{ed}^d}\tensor{k}{_{(a}_{b)}^e} - \tensor{k}{_e_{(a} ^d}\tensor{k}{_{b)}_d^e} &=&2F(t, t')^2\delta^{(3)}_{ab}.
\end{eqnarray}
Combining these results yields
\begin{equation}\label{intermedio}
H_{ab}= \delta^{(3)}_{ab}\left[\frac{\partial F}{\partial t} + 2F^2\right] - 3u_a u_b\frac{\partial F}{\partial t}.
\end{equation}
Repeating the calculation at $x'$ and applying the parallel transporters produces
\begin{align}
\frac{\partial F}{\partial t} + \frac{\partial F}{\partial t'} + F^2 &= -8 \pi P.\\\label{ecuacion de movimiento F}
3 F^2 &= 8 \pi \rho,
\end{align}
Importantly, note that the second relation is algebraic.

To solve this system, it is convenient to introduce 
\begin{equation}
\delta t = t + t', \qquad
\Delta t = t - t',
\end{equation}
and define
\begin{equation}
F(t,t') = \tilde{F}(\Delta t, \delta t).
\end{equation}
Introducing the algebraic equation into Eq.~\eqref{ecuacion de movimiento F}, it becomes
\begin{equation}\label{ecuaciondiferencialG}
\frac{\partial \tilde{F}}{\partial \delta t} = -\frac{8 \pi}{3}(\rho + 3P),
\end{equation}
which is subject to
\begin{equation}
\label{ecuacionalgebraicaG}
3\tilde{F}^2 = 8 \pi\rho.
\end{equation}

To proceed, we must specify an equation of state. Our proposal is to adopt the simplest alternative, directly analogous to what is done in standard cosmology: assume that the bitensorial pressure is proportional to the bitensorial density. It is important to emphasize, however, that the physical interpretation of these bitensorial quantities remains unclear. Furthermore, they represent only a small correction to the total density and pressure, which continue to be dominated by their classical components. Hence, they are not restricted to be nonnegative.

Assuming an equation of state of the form $P = \omega \rho$, with $\omega$ a real constant, and using Eq.~\eqref{ecuacionalgebraicaG}, Eq.~\eqref{ecuaciondiferencialG} becomes
\begin{equation}\label{ecuaciondefinitivarho}
\frac{\partial \rho}{\partial \delta t} = -\sqrt{\frac{32 \pi}{3}}\,(1+3\omega)\,\rho^{3/2}.
\end{equation}

For concreteness, we restrict from this point onwards to Hadamard states. In this case, in the coincidence limit~\cite{JuarezDeSitter}, $\rho$ goes as $(\Delta t)^{-4}$. However, from \eqref{ecuaciondefinitivarho}, one can show that, for $\rho$ to diverge in the coincidence limit, it is necessary that $\omega = -1/3$. This, in turn, implies that $\rho$ depends only on $\Delta t$. Consequently,
\begin{equation}\label{bisolucionmodelo}
\tensor{k}{_{abc}} = \frac{2\tau}{(t-t')^2}\, u_{[a}\, \delta^{(3)}_{c]b}.
\end{equation}
where $\tau$ is a constant with units of time. Observe that this expression is symmetric under $t \leftrightarrow t'$.

At this point the model is formally solved: for any given homogeneous and isotropic matter content in a Hadamard state and any FRW background metric, the bitensorial connection satisfying the field equations is completely determined. The remaining task is to extract the physical information encoded in this object.

\subsection{Effective geometrical description}

From a mathematical standpoint, the model has been solved under the relevant assumptions. However, its physical implications remain to be extracted. In quantum theories, when certain degrees of freedom are ignored, one traces over them~\cite{NielsenChuang}. In this spirit, our proposal for obtaining local physical quantities at \(x\) is to treat the dependence on \(x'\) effectively by integrating over it (restricted to a convex normal neighborhood).

The precise prescription we propose is as follows. Our goal is to extract an effective contorsion that is an ordinary tensor, $\tensor{\bar{k}}{_a_b^c}(x)$. Moreover, inspired by Bell’s theorem, the type of nonlocalities we aim to incorporate operate outside the light cone. Thus, for any point $x$ we integrate over all $x' \in \mathcal{C}$, where $\mathcal{C}$ denotes the set of all points that are spatially separated from $x$ and lie within a normal convex neighborhood. The integration takes the form
\begin{equation}\label{abc}
\tensor{\bar{k}}{_a_b^c}(x)
    = \frac{1}{V_{\mathcal{C}}}
      \int_{\mathcal{C}}
      \tensor{k}{_a_b^c}(x,x') \sqrt{-g}\, \text{d}^4x',
\end{equation}
where $\sqrt{-g}\, \text{d}^4x'$ is the appropriate integration element and $V_{\mathcal{C}}$ is the $4$-volume of $\mathcal{C}$, introduced for dimensional consistency. Note that this integral is well defined because $ \tensor{k}{_a_b^c}(x,x')$ is a scalar with respect to $x'$.

Moreover, under the symmetries at hand and within the level of approximation we are working, Eq.~\eqref{abc} reduces to integrating $\tau/(t-t')^2$. In spherical coordinates, this becomes
\begin{equation}
\int_{\mathcal{C}} \frac{\tau}{(t-t')^2} \sqrt{-g}\, \text{d}^4 x'
= 4\pi \tau \int_{\mathcal{C}} \frac{r'^2}{(t - t')^2}\, \text{d}t'\, \text{d}r',
\end{equation}
where the angular coordinates integrate trivially. Since the integration is performed outside the light cone, the integration limits are not independent. In fact, the light-cone condition restricts the limits of the $t'$ integral to the interval $(-r', r')$. Moreover, given that we also work within a normal convex neighborhood, we assume that $r'$ extends up to a cutoff, $r_c>0$. Integrating over this region and normalizing with the volume yields
\begin{equation}
\tau	\frac{\int_0^{r_c} \int_{-r'}^{r'} \frac{r'^2}{(t - t')^2}\, \text{d}t'\, \text{d}r'}{\int_0^{r_c} \int_{-r'}^{r'} r'^2\, \text{d}t'\, \text{d}r'} =\phi(t)
\end{equation}
where we define
    \begin{equation}
  \phi(t)  =-\frac{2\tau}{r_c^2}\left(\frac{1}{2}+\frac{t^2}{r_c^2}\ln\left[1-\frac{r_c^2}{t^2}\right]\right) .\label{phi}
	 \end{equation}
Thus, the effective contorsion becomes
\begin{equation}
\tensor{\bar{k}}{_a_b_c}(x)= 2\phi(t) u_{[a}\, \delta^{(3)}_{c]b}.
\end{equation}

We want to extract an effective cosmological constant. A natural approach is to compute the Ricci curvature associated with $\tensor{\bar{k}}{_a_b^c}$, denoted $R_{ab}(\bar{k})$, and compare it with the Ricci tensor of the FRW metric, $R_{ab}(a)$. The only nonzero components are the time--time component and those with identical spatial indices. Comparing these components as obtained by the two methods yields
\begin{align}
    -3\frac{\ddot{a}}{a} &= -3\dot{\phi}, \\
    2\phi^2 + \dot{\phi} &= \ddot{a}a + 2\dot{a}^2.
\end{align}
Since we are only interested in the late-time universe, we work within the approximation in which the time variation of $\phi$ is negligible. Thus, we take $\phi(t)\approx \phi(t_0)$ and find
\begin{equation}
    \phi(t_0)^2 = \dot{a}(t_0)^2,
\end{equation}
or equivalently,
\begin{equation}\label{logaritmo}
    H_0 = \frac{2}{r_c^2}\left|
        \frac{\tau}{2}
        + \frac{\tau t_0^2}{r_c^2}
        \ln\!\left(1 - \frac{r_c^2}{t_0^2}\right)
    \right|.
\end{equation}
Note that this expression represents a strictly positive additional contribution to the classical cosmological constant that turns on in the late universe. Hence, it may offer an explanation for the \(H_0\) tension~\cite{H0tension}. For a complete assessment, however, the predictions of the model in the early universe must also be considered, a task we leave for future work.

Interestingly, Eq.~\eqref{logaritmo} shows that the analysis forbids $r_c$ from being equal to $t_0$. This suggests that $r_c$ must lie below the Hubble radius. To obtain a simpler expression useful for constraining the free parameter, we consider the regime $r_c \ll t_0$. To leading order, Eq.~\eqref{logaritmo} becomes
\begin{equation}\label{resultadocosmologia}
    H_0 = \frac{|\tau|}{r_c^2}.
\end{equation}
Using the observed value $H_0 \approx 10^{-18}\,\mathrm{s}^{-1}$, assuming that classical matter fields do not contribute to the cosmological constant, and taking into account the experimentally established fact~\cite{Entrelazamiento, Entrelazamiento2} that quantum entanglement has been produced over distances $r_c > 10^6\,\mathrm{m}$, we obtain the bound
\begin{equation}
    |\tau| > 10^{-6}\,\mathrm{s},
\end{equation}
which is extremely small when compared with cosmological time scales. We now turn to the effects of the model in the Newtonian limit.

\section{Newtonian limit}

One of the most pressing questions in modern physics is: How does matter in a quantum state gravitate? An illustrative example is a source prepared in a spatial superposition. Analogous questions naturally arise for superpositions associated with other properties, such as energy or momentum. Fortunately, the bitensorial model proposed here provides a framework that can address these issues and extends beyond the scope of semiclassical gravity.

Throughout this analysis we assume that spacetime is nearly flat, so that the linearized Einstein equations hold. Hence, we work to first order in \(\epsilon\), under the assumption that the bitensorial connection is itself of order \(\epsilon\). In addition, the Newtonian limit requires a nonrelativistic approximation in which all relative velocities are much smaller than the speed of light. This can be effectively implemented by expanding around the formal limit in which the speed of light is taken to be infinite.

The first step in this analysis is to identify an appropriate bitensorial description of the matter fields. To address this point we can rely on insights from conventional Quantum Field Theory, as is done next.

\subsection{Matter description}

We begin this subsection by constructing the two-point function for a real, massive Klein--Gordon field, which we use to model matter. The smeared field operator can be written as
\begin{equation}
\phi(x)=\int \text{d}^3p \left(f^*(\vec{p}) e^{-ip\cdot x} a^{\dagger}_{\vec{p}} + f(\vec{p}) e^{ip\cdot x} a_{\vec{p}}\right),
\end{equation}
where \(f(\vec{p})\) is the smearing function. Here the star denotes complex conjugation, and \(p^a\) is the on-shell four-momentum with spatial components \(\vec{p}\). The operators \(a^{\dagger}_{\vec{p}}\) and \(a_{\vec{p}}\) are the creation and annihilation operators, respectively, associated with \(\vec{p}\). We note that, in standard Quantum Field Theory, the form of \(f(\vec{p})\) is irrelevant since scattering amplitudes are extracted via the LSZ reduction formula~\cite{LSZ}, which relies on asymptotic states. In contrast, the present work focuses on finite-time, localized quantum configurations, where the choice of smearing function can play a nontrivial role.

The corresponding two-point function is
\begin{equation}
G(x,x')= 
\bra{0}\phi(x)\phi(x')\ket{0}
=\chi(x)\chi^{\dagger}(x')+\chi(x')\chi^{\dagger}(x)+H(x,x'),
\end{equation}
where \(H(x,x')\) denotes the divergent contribution, which is typically removed by normal ordering, and
\begin{equation}
\chi(x)=\int \text{d}^3p\, f(\vec{p})\, e^{-ip\cdot x}.
\end{equation}
For the purposes of the present calculation, we take \(f(\vec{p})\) to be real, purely for convenience, and we neglect the term \(H(x,x')\), as is commonly done. It is worth recalling, however, that \(H(x,x')\) does play a role, as it effectively encodes a boundary condition.

The energy--momentum bitensor, given by the point-splitting prescription, is
\begin{equation}
\braket{T_{ab'}}=\partial_a\partial_{b'}G(x,x')
-\frac{1}{2}\eta_{ab'}\left(\tensor{g}{_{c'}^{c}}\partial_c\partial^{c'}+m^2\right)G(x,x').
\end{equation}
This expression follows from the classical energy–momentum tensor, with the squared field replaced by $G(x,x')$. Note that, given the bitensorial structure, each derivative acts on a quantity that behaves as a scalar.

For concreteness, we adopt a Gaussian smearing function \(f(\vec{p})\), which yields Gaussian spatial distributions. We also focus on a static configuration in which the particle acting as the gravitational source is in a superposition of being centered around the spatial points \((0,0,a)\), with amplitude \(\alpha\), and \((0,0,-a)\), with amplitude \(\beta=\sqrt{1-\alpha^2}\) (again, the amplitudes are taken to be real, for simplicity).

Clearly, the setup exhibits cylindrical symmetry around the $z$-axis. Thus, it is convenient to introduce the distance to this axis, $\rho$, which satisfies $\rho^{2} = x^{2} + y^{2}$ (not to be confused with the density introduced in Sec.~\ref{Sec:Cosmology}). Under these assumptions we can write
\begin{equation}\label{gaussianas}
	\chi(\vec{x})=
	\frac{\chi_z(z)}{(m\sigma^3)^{1/2} (2\pi)^{3/2}}
	\exp\!\left(-\frac{1}{2}\frac{\rho^2}{\sigma_\perp^2}\right),
\end{equation}
where
\begin{equation}\label{partezchi}
	\chi_z(z) = \alpha\, \exp\!\left[-\frac{(z+a)^2}{2\sigma_z^2}\right]
	+\beta\, \exp\!\left[-\frac{(z-a)^2}{2\sigma_z^2}\right].
\end{equation}
Here $\sigma_x=\sigma_\perp=\sigma_y$ and \(\sigma^3 = \sigma_\perp^2 \sigma_z\) denotes the product of the Gaussian widths.

In the Newtonian limit, the dominant component of the energy--momentum tensor is the time--time component. In the static case this component is further dominated by
\begin{equation}\label{T00NR}
\braket{T_{00'}} \approx \frac{m^2}{2}\, G(x,x').
\end{equation}
This is the gravitational source we use to study Newtonian limit of the model.

\subsection{Bitensorial Einstein equation}

Since the matter source is static, the geometrical side of Eq.~\eqref{esquema2} is also dominated by the time--time component, which takes the form
\begin{equation}\label{ecuaciondelmodelo2}
\delta^b_{0'}\,\eta_{0b}\eta^{de}
\left[\partial_c\tensor{k}{_d_e^c}-\partial_e\tensor{k}{_d_c^c}\right]
+ \delta^{a'}_{0}\,\eta_{a'0'}\eta^{d'e'}
\left[\partial_{c'}\tensor{k}{_{d'}_{e'}^{c'}}-\partial_{e'}\tensor{k}{_{d'}_{c'}^{c'}}\right].
\end{equation}
Using the index symmetries of the contorsion, the spatial components can be conveniently written as
\begin{equation}\label{Kespacial}
\tensor{k}{_i_j_k}(x,x')=\tensor{\epsilon}{_i_k^\ell}\,v_{j\ell}(x,x'),
\end{equation}
where \(\tensor{\epsilon}{_i_k^\ell}\) is the volume element of three-dimensional Euclidean space, and the indices are raised with the Euclidean metric \(\delta_{j\ell}=\mathrm{diag}(1,1,1)\). In this expression, \(v_{j\ell}\) is a bitensor of type \((0,2)\times(0,0)\), and its antisymmetric part, which is the only component that contributes to the present calculation, is denoted by \(v^{A}_{j\ell}\).

It is now straightforward to find the relevant bitensorial equation. A direct computation shows that substituting the ansatz~\eqref{Kespacial} into Eq.~\eqref{ecuaciondelmodelo2}, and equating it to Eq.~\eqref{T00NR}, yields
\begin{align}\label{ecuacionaresolverBMV}
	\epsilon^{kjl}\partial_k v_{jl}^A
	+\epsilon^{k'j'l'}\partial_{k'}v^A_{j'l'}
	=4\pi m^2\, \chi(\vec{x})\chi(\vec{x}').
\end{align}
This equation may be rewritten as
\begin{align}\label{ecuacionprincipal2}
\partial_x B + \partial_y C + \partial_z A
+ \partial_{x'} B + \partial_{y'} C + \partial_{z'} A
= 2\pi m^2 \chi(\vec{x})\chi(\vec{x}'),
\end{align}
where the independent components of \(v^{A}_{j\ell}\), in cylindrical symmetry, are defined as
\begin{equation}\label{ABC}
v^{A}_{xy}=A(\vec{x},\vec{x}'),\quad
v^{A}_{yz}=B(\vec{x},\vec{x}')=v^{A}_{zx}.
\end{equation}
What is more, this symmetry allows us to write
\begin{equation}
\chi(\vec{x})
=\frac{\chi_z(z)}{(m\sigma^3)^{1/2}(2\pi)^{3/2}}\,
\, \exp\!\left(-\frac{\rho^2}{2\sigma_\perp^2}\right).
\end{equation}

At this stage it is important to recall that the model is underdetermined. This manifests itself here as having more unknowns than equations. To proceed, and to extract physical insight from this still incomplete yet valuable model, we impose $B=0$. Under these considerations, the equation to solve becomes
\begin{align}\label{ecuacionprincipal3}
\partial_z A(\vec{x},\vec{x}') + \partial_{z'} A(\vec{x},\vec{x}') =
\frac{   m \chi_z(z)\chi_z(z')}{(2\pi)^2\sigma^3}
\exp\!\left(-\frac{\rho^2+\rho'^2}{2\sigma_\perp^2}\right).
\end{align}

The ansatz
\begin{equation}\nonumber
A(\vec{x},\vec{x}') = A_1(z)\, A_2(z')\, A_3(\rho,\rho'),
\end{equation}
leads to
\begin{equation}
A_3(\rho,\rho')\left[A_2(z')\frac{\text{d}A_1(z)}{\text{d}z}
+ A_1(z)\frac{\text{d}A_2(z')}{\text{d}z'}\right]
= \frac{m\,\chi_z(z)\chi_z(z')}{(2\pi)^2\sigma^3}
\exp\!\left(-\frac{\rho^2+\rho'^2}{2\sigma_\perp^2}\right).
\end{equation}
This expression allow us to conclude that
\begin{equation}
A_3(\rho,\rho') =
\frac{m}{(2\pi)^2\sigma^3}
\exp\!\left(-\frac{\rho^2+\rho'^2}{2\sigma_\perp^2}\right).
\end{equation}
The remaining equation to solve is
\begin{equation}\nonumber
A_2(z')\frac{\text{d}A_1(z)}{\text{d}z}
+ A_1(z)\frac{\text{d}A_2(z')}{\text{d}z'} =
\chi_z(z)\chi_z(z'),
\end{equation}
which can be rewritten in terms of \(A_4(h)\), with \(h = z + z'\), as
\begin{equation}\nonumber
\frac{\text{d}A_4(h)}{\text{d}h}
= \chi_z(h - z')\, \chi_z(h - z).
\end{equation}
Integrating this expression yields
\begin{align}\nonumber
A_4(z,z') =&\,
\sigma_z\frac{\sqrt{\pi}}{2}\,
\exp\!\left[-\frac{(z - z')^2}{4\sigma_z^2}\right]
\Bigg[
\alpha^2 \Erf\!\left(\frac{a + (z + z')/2}{\sigma_z}\right)
- \beta^2 \Erf\!\left(\frac{a - (z + z')/2}{\sigma_z}\right)
\\\nonumber
&\quad
+ 2\alpha\beta\,
\exp\!\left(-\frac{a^2}{\sigma_z^2}\right)
\cosh\!\left(\frac{a(z - z')}{\sigma_z^2}\right)
\Erf\!\left(\frac{z + z'}{2\sigma_z}\right)
\Bigg],
\end{align}
where we use the ``error function''
\begin{equation}
\Erf(z) = \frac{2}{\sqrt{\pi}}\int_0^z \exp\!\left(-t^2\right)\,\text{d}t.
\end{equation}
Collecting all the terms, the solution of the model reads
\begin{equation}\label{solucionmodelo}
\tensor{k}{_i_j^k}(x,x') = A(\vec{x},\vec{x}')
\left[
-\delta^y_i\delta^y_j\delta^k_z
+ \delta^z_i\delta^y_j\delta^k_y
+ \delta^z_i\delta^x_j\delta^k_x
- \delta^x_i\delta^x_j\delta^k_z
\right],
\end{equation}
where
\begin{align}\nonumber
A(\vec{x},\vec{x}') =&\,
\frac{m\,\sigma_z\sqrt{\pi}}{2(2\pi)^2\sigma^3}
\exp\!\left[-\frac{\rho^2+\rho'^2}{2\sigma_\perp^2}\right]
\exp\!\left[-\frac{(z - z')^2}{4\sigma_z^2}\right]
\\
&\times
\Bigg[
\alpha^2 \Erf\!\left(\frac{a + (z + z')/2}{\sigma_z}\right)
- \beta^2 \Erf\!\left(\frac{a - (z + z')/2}{\sigma_z}\right)
\nonumber\\
&+ 2\alpha\beta\, \exp\!\left(-\frac{a^2}{\sigma_z^2}\right)
\cosh\!\left(\frac{a(z - z')}{\sigma_z^2}\right)
\Erf\!\left(\frac{z + z'}{2\sigma_z}\right)
\Bigg].
\end{align}

Thus far, we have obtained the bitensorial connection that is a solution of the experimental configuration under consideration. In the next subsection, we construct the corresponding effective tensorial quantities.

\subsection{Physical implications}

As in the cosmological case, the effective connection is obtained by integrating the bitensorial connection, given in Eq.~\eqref{solucionmodelo}, over the primed coordinates. However, in the nonrelativistic setting there is no light cone: it collapses into the Newtonian instantaneous surface. Consequently, the effective connection is given by
\begin{equation}
\tensor{\bar{k}}{_i_j^k}(\vec{x})
= \frac{1}{L^3}\int_{\mathbb{R}^3}
\tensor{k}{_i_j^k}(\vec{x},\vec{x}')\, \text{d}^3x',
\end{equation}
where the relevant integration domain is $\mathbb{R}^3$, which is a normal convex neighborhood (a curvature cutoff would be of order $\epsilon$ and is therefore negligible at the present level of approximation). The factor $L^3$, corresponding to the volume of the integration region, is included to ensure the correct units.

The relevant quantity to be integrated is $A(\vec{x},\vec{x}')$. We define
\begin{equation}
\bar{A}(\vec{x}) = \frac{1}{L^{3}} \int A(\vec{x},\vec{x}')\,\mathrm{d}^{3}x' 
= \bar{A}_z(z)\,\bar{A}_\perp(\rho),
\end{equation}
where
\begin{align}\nonumber
\bar{A}_z(z)=&\,\frac{1}{L}\int 
\exp\!\left[-\frac{(z-z')^{2}}{4\sigma_z^{2}}\right]
\Bigg[
\alpha^{2}\,\Erf\!\left(\frac{a+\frac{z+z'}{2}}{\sigma_z}\right)
-\beta^{2}\,\Erf\!\left(\frac{a-\frac{z+z'}{2}}{\sigma_z}\right)\\\nonumber
&\quad+2\alpha\beta\,\exp\!\left(-\frac{a^{2}}{\sigma_z^{2}}\right)
\cosh\!\left(\frac{a(z-z')}{\sigma_z^{2}}\right)
\Erf\!\left(\frac{z+z'}{2\sigma_z}\right)
\Bigg]\,\mathrm{d}z'\\
=&\frac{\pi\sigma_z}{L}\!\left[
\Erf\!\left(\frac{z+a}{\sqrt{2}\sigma_z}\right)(\alpha^{2}+\alpha\beta)
+\Erf\!\left(\frac{z-a}{\sqrt{2}\sigma_z}\right)(\beta^{2}+\alpha\beta)
\right].
\end{align}
Note that, in the last step, we use the identity
\begin{equation}
\int_{-\infty}^{\infty}\Erf(z')\,\exp\!\left[-(a z'+b)^2\right]\mathrm{d}z'
= -\frac{\sqrt{\pi}}{a}\,
\Erf\!\left(\frac{b}{\sqrt{a^{2}+1}}\right).
\end{equation}
Moreover, we have
\begin{equation}
\bar{A}_\perp(\rho)
=\frac{m\sigma_z\sqrt{\pi}}{2L^{2} \sigma^{3}(2\pi)^{2}}
\int \,
\exp\!\left(-\frac{\rho^{2}+\rho'^{2}}{2\sigma_\perp^{2}}\right)
\rho'\,\mathrm{d}\rho'\,\mathrm{d}\theta' =\frac{m}{2\pi L^2}\,
\exp\!\left(-\frac{\rho^{2}}{2\sigma_\perp^{2}}\right).
\end{equation}
Combining these results yields
\begin{equation}
\bar{A}(\vec{x})
=\frac{m \sigma_z}{2L^{3}}
\left[
\Erf\!\left(\frac{z+a}{\sqrt{2}\sigma_z}\right)(\alpha^{2}+\alpha\beta)
+\Erf\!\left(\frac{z-a}{\sqrt{2}\sigma_z}\right)(\beta^{2}+\alpha\beta)
\right]
\exp\!\left(-\frac{\rho^{2}}{2\sigma_\perp^{2}}\right),
\end{equation}
which, together with Eq.~\eqref{solucionmodelo}, gives the effective connection.

With this, we can find the force due to the quantum effects, which acts on top of that generated by the metric. We denote the components of this force by $F^\mu$, and define it through
\begin{equation}\label{geodesica3}
u^\nu\mathring{\nabla}_\nu u^\mu=\frac{F^\mu}{M},
\end{equation}
where $u^a$ is the tangent of the trajectory of a test particle of mass $M$ and $\mathring{\nabla}_\nu$ is the derivative associated with the purely metric connection (to first order in $\epsilon$). We can readily show that
\begin{equation}\label{geodesica4}
F^\mu= M \tensor{\bar{k}}{_\nu_\rho^\mu}u^\nu u^\rho.
\end{equation}
Thus, the temporal component of the effective force vanishes, as expected in the Newtonian limit, and the spatial components take the form
\begin{equation}\label{fz}
F^i=2M\bar{A}(\vec{x})\,v^z v^i,\end{equation}
and
\begin{equation}
F^z=-M\bar{A}(\vec{x})v_\perp^2,
\end{equation}
where $i=x,y$, $\vec{v}$ is the $3$-velocity of the test particle (which coincides with the initial $3$-velocity to first order in $\epsilon$), and $v_\perp^2=(v^x)^2+(v^y)^2$. Observe that the effective force is quadratic on the test particle’s velocity. This behavior is not entirely new: it also appears when considering gravitoelectromagnetic phenomena~\cite{MashhoonGM}. However, this is a clear indication that this effective force is not conservative. 

Another notable feature is that the force components are proportional to the Gaussian width in the \(z\) direction, \(\sigma_z\). Consequently, as \(\sigma_z \rightarrow 0\), the effective force vanishes. In other words, in the most classical configuration of the source, the effective force disappears. In this limit, the function given in Eq.~\eqref{gaussianas} reduces to a sum of Dirac deltas. Still, to consider a physically realistic configuration, one needs to assume that the particle is confined in a harmonic trap~\cite{Bosetrampasarmonicas}, for which the dispersion satisfies
\begin{equation}
\sigma_z = \sqrt{\frac{\hbar}{m\omega_z}}.
\end{equation}
Thus, when \(\hbar \to 0\), \(\sigma_z\) vanishes, and so does the effective force.

Next, we address whether the effective force is attractive, as expected for a gravitational effect. The simplest way to verify this is to turn off one of the packets by setting, say, \(\alpha = 0\). For a test particle traveling in the \(z = 0\) plane, when it is at the origin, it is straightforward to see that \(F^x =0= F^y\). Thus, the effective force becomes
\begin{equation}
F^z = \frac{\sigma_z m M v_\perp^2}{2L^3}\, \Erf\!\left(\frac{a}{\sqrt{2}}\right),
\end{equation}
which is positive, indicating that the force points toward the packet. This calculation can be generalized to the case where both packets are present. In this situation, the effective force on the test particle still behaves as an attractive one, with a factor \(\alpha^2 - \beta^2\) determining the force direction.

An interesting limit is \(\alpha = \beta\). This case is clearly invariant under \(z \rightarrow -z\). Indeed, since the error function is odd, the effective force satisfies \(\bar{A}(x,y,-z) = -\bar{A}(x,y,z)\). Also, the effective force vanishes at the origin for all particles traveling in the \(z = 0\) plane, as expected. 

In the far region, \(z \gg a\), \(F^z\) becomes independent of \(\alpha\) and \(\beta\), meaning that the test particle experiences the total gravitational influence of the superposed source. Finally, for \(a \ll \sigma_z\) and \(z \ll \sigma_z\), a Taylor expansion of the error function yields
\begin{equation}
F^z = -\frac{\sigma_z m M v^2}{\sqrt{2\pi}L^3}
\exp\!\left(-\frac{\rho^2}{2\sigma_\perp^2}\right)
\left[\frac{z}{\sigma_z} + \frac{a}{\sigma_z}\left(\alpha^2 - \beta^2\right)\right],
\end{equation}
which, at least near \(\rho = 0\), resembles the force of a simple harmonic oscillator plus a constant shift toward the heavier packet. These properties demonstrate that the model reproduces all the expected symmetries of the setup, offering a nontrivial check of its internal consistency. In addition, the model predicts a definite effective gravitational force exerted by a particle in a spatial superposition on a test particle; a concrete effect that may, in principle, be accessible to experimental scrutiny.
 
\section{Conclusions}

In this work, inspired by Bell’s inequalities and by the structure of the expectation value of the energy--momentum tensor in Quantum Field Theory, we have proposed a bitensorial extension of GR. The bitensorial aspects of the model are incorporated directly into the connection, yielding a purely geometric yet intrinsically nonlocal theory. Notably, under the assumptions of homogeneity and isotropy, and in the Newtonian limit, the model becomes analytically solvable.

In the cosmological context, the late-time solution produces a positive contribution to the cosmological constant and reveals a relation between geometric quantities, such as the ``size'' of a normal convex neighborhood, and quantum correlations. The model is expected to be particularly relevant in high-curvature regimes, including the early universe. Once the early-universe analysis is completed, it may be possible to assess the model’s implications for current cosmological tensions, such as the one involving $H_0$.

We have also derived an effective force within the Newtonian limit. This force depends on the test particle's velocity and is inherently nonconservative. Importantly, it is, in principle, amenable to empirical testing through precise measurements of deviations from standard Newtonian trajectories in laboratory-scale experiments.

Perhaps the central message of this work is that new and testable predictions can arise from ``minimal'' modifications to our existing theoretical framework, particularly when these modifications are guided by empirical insights such as the violation of Bell’s inequalities. In this way, the model brings GR and quantum mechanics closer together by incorporating features that have not typically been considered in gravitational physics. This example illustrates that it is not necessary to have a complete quantum theory of gravity to look for novel observable effects emerging from the interplay between gravitational and quantum phenomena. If any of the effects predicted by models of this kind were eventually observed, they would provide valuable guidance for future attempts to reconcile the two theories.

Finally, this framework offers a new perspective on the measurement problem. In standard semiclassical gravity, the collapse of a quantum state induces a discontinuity in the spacetime metric~\cite{Adolfo,Diez}. In contrast, within our bitensorial formulation, the collapse leads to a discontinuity in the bitensorial connection, whose physical implications remain to be explored.

\section*{Acknowledgments}
We appreciate valuable feedback from E. Ayón-Beato, B. Arderucio, C. Chryssomalakos, B.A. Juárez-Aubry, S. Modak, E. Okon, and D. Sudarsky. This work was supported by the UNAM-DGAPA-PAPIIT grant IN101724 and by the SECIHTI through its graduate school scholarship program.

\end{document}